\def \be {\begin{equation}}
\def \ee {\end{equation}}
\def \bea {\begin{eqnarray}}
\def \eea {\end{eqnarray}}
\def \nn {\nonumber}

\def \del {\partial}
\def \dels {\partial\kern-.5em / \kern.5em}
\def \As {{A\kern-.5em / \kern.5em}}
\def \Ds {D\kern-.7em / \kern.5em}

\def \a {\alpha}
\def \b {\beta}

\def \d {\delta}
\def \eps {\epsilon}

\def \s {\sigma}

\def \Om {\Omega}

\def \th {\theta}

\def \II {I\hspace{-.1em}I\hspace{.1em}}

\def \F {{\cal F}}

\def \db {{\bf d}}

\documentclass[12pt]{article}

\begin{document}
\begin{titlepage}
%\catcode`\@=11
%\catcode`\@=12
%\twocolumn[\hsize\textwidth\columnwidth\hsize\csname%
%@twocolumnfalse\endcsname

%\draft
\begin{center}
\hfill hep-th/0005159\\
\vskip .5in

\textbf{\Large \textbf{Noncommutative D-Brane
in Non-Constant \\ NS-NS B Field Background}}

\vskip .5in
{\large Pei-Ming Ho, Yu-Ting Yeh}
\vskip 15pt

{\small \em Department of Physics, National Taiwan
University, Taipei 106, Taiwan, R.O.C.}

\vskip .2in
\sffamily{
pmho@phys.ntu.edu.tw\\
r7222049@ms.cc.ntu.edu.tw}

\vspace{60pt}
%\maketitle
\end{center}
\begin{abstract}

We show that when the field strength H
of the NS-NS B field does not vanish,
the coordinates x and momenta p of an open string endpoints
satisfy a set of mixed commutation relations among themselves.
Identifying x and p with the coordinates and derivatives
of the D-brane world volume,
we find a new type of noncommutative spaces
which is very different from those
associated with a constant B field background.

\end{abstract}
%\pacs{PACS numbers: 11.25.-w, 11.25.Mj, 11.25.Sq}%]
\end{titlepage}
%\begin{narrowtext}
\setcounter{footnote}{0}

\newpage

\section{Introduction}

In the past few years there has been a growth
in the interest in noncommutative geometry,
which appears in string theory in several different ways.
To our knowledge the first paper on this topic is \cite{Witten}.
For the earlier focus on the use of noncommutative geometry
in matrix theory compactifications, see, for instance, \cite{NC}.
In this paper we follow \cite{CH1,CH2} and
find a new type of noncommutative spaces
that appear naturally in string theory
as a description of the D-brane worldvolume.

In \cite{CDS,DH}, it was proposed that
the matrix theory compactified on a torus with
constant 3-form C field background should
be described by a field theory living on
a noncommutative space whose coordinates
satisfy a noncommutative algebra of the form
\be \label{th}
[x^i,x^j]=i\th^{ij},
\ee
where $\th^{ij}=RC^{-ij}$ and
$R$ is the light cone radius of $X^{-}$.
As an evidence,
the BPS spectrum on the quantum torus was given in \cite{Ho,BMZ},
and this conjecture was later derived \cite{CHL}
from the discrete light cone quantization of the membrane action.
Via string dualities, it follows that
in the background of a constant NS-NS $B$ field,
the low energy field theory of a flat D-brane in flat spacetime
lives on a noncommutative space described by (\ref{th}) where
\bea \label{thMB}
&\th^{ij}=-2\pi\a'(G^{-1}BM^{-1})^{ij}, \\
&M_{ij}=G_{ij}-B_{ik}G^{kl}B_{lj},
\eea
where $G_{ij}$ is the spacetime metric viewed by closed strings.
Here we assumed that the $U(1)$ field strength $F=dA$ vanishes.
In general, since $\F=B-F$ is the gauge invariant quantity,
it is natural to replace $B$ by $\F$ in (\ref{thMB}).
\footnote{For the relation between different noncommutativity
due to different choices of background values, see \cite{SW}.}
The simplest way to derive this result is
to quantize an open string ending on the D-brane \cite{CH1,CH2,ASS}.
This serves as a direct evidence for
the noncommutativity of D-brane worldvolume
in the B field background.
Later it was shown \cite{BS,Yin} that
for the sake of deriving endpoint commutation relations,
it is sufficient to approximate the open string by
a straight line stretched between its endpoints.
This is equivalent to say that we quantize the open string
in the low energy limit ($\a'\rightarrow 0$).
Other approaches for calculating
the D-brane worldvolume noncommutativity
can be found in e.g. \cite{Schom,G-C,Alek}.

In this paper we consider the more general case
of a curved D-brane in a curved spacetime with a nonconstant $B$ field.
Obviously, eq.(\ref{thMB}) will not continue to hold,
because eq.(\ref{th}) may not satisfy the Jacobi identities anymore.
We will show that in the generic case
$\th$ will be replaced by a function depending
not only on the coordinates $x$ but also on the derivatives $\del$.
The D-brane worldvolume thus belongs to a new type of noncommutative spaces
which are described by a mixed algebra of $x$ and $\del$.

\section{Generic Case}

The bosonic part of the action for an open string ending on a D-brane
in the background of a NS-NS B field is
\footnote{
We have absorbed the dilaton factor in $G_{\mu\nu}$ and $\F_{ij}$.}
\begin{equation} \label{action}
S_{B}=\int d\tau L=\frac{1}{4\pi \alpha ^{\prime }}\int d^{2}\sigma
\left[\eta^{\a\b}G_{\mu\nu}\partial_{\a}X^{\mu}\partial_{\b}X^{\nu}
+\eps^{\a\b}\F_{ij}\partial_{\a}X^{i}\partial_{\b}X^{j}\right],
\end{equation}
where $\eta^{\a\b}=\mbox{diag}(1,-1)$, and $\eps^{01}=-\eps^{10}=1$.
We use $X^{i}$ and $X^{a}$ to denote longitudinal and transverse
directions for the D-brane, respectively,
and use $X^{\mu}$ for all spacetime directions.
For simplicity we assume that $\F_{a\mu}=0$, $G_{ia}=0$,
and that $\F_{ij}$ is invertible.

The conjugate momentum of $X^{\mu}$ is
\be \label{P}
P_{\mu}=\frac{1}{2\pi\a'}\left[G_{\mu\nu}\dot{X}^{\nu}
+\F_{\mu i}X^{'i}\right],
\ee
and the boundary conditions are
\be \label{BC}
G_{ij}X^{'j}+\F_{ij}\dot{X}^{j}=0, \quad
\dot{X}^{a}=0.
\ee

In the limit $\a'\rightarrow 0$,
the oscillation modes can be ignored since
their energies are proportional to $1/\a'$.
The bulk of the string is now determined by its boundary.
In principle, one can try to solve the wave equations for $X^{\mu}$
and pick out the lowest energy mode which survives
the limit $\a'\rightarrow 0$.
Here we avoid the complexity by focusing on the low energy limit
in which both $\dot{X}$ and $X'$ are very small.
This means that the string is very short and moves very slowly,
so that the spacetime appears to be almost flat
and $\F$ is almost constant.
Therefore we can use the results in \cite{CH1} for a flat background
and see that eq.(\ref{BC}) holds for all $\s$ for the lowest energy mode.
It follows that $P_a=0$ and
\be \label{Pi}
P_i=\frac{1}{2\pi\a'}
    \left(\F_{il}-G_{ij}\F^{jk}G_{kl}\right)X^{'l},
\ee
where $\F^{ij}$ is the inverse matrix of $\F_{ij}$.

Since $P_a$ vanishes, $X^a$ will be just constant
for the whole string, and so we will ignore them from now on.
Let
\be
\hat{\F}=\frac{1}{2\pi\a'}(\F-G\F^{-1}G),
\ee
then a shorthand of (\ref{Pi}) is
\be
P=\hat{\F}X', \label{Pi1} \\
\ee

The symplectic two-form which determines
the commutation relations among $X$ and $P$ is
\be
\Om=\int d\s \left(\db X^i \db P_i\right).
\ee
Using (\ref{Pi1}), and the identity $dF=0$, we find
\be \label{Om}
\Om=\frac{1}{2}\left[\db X^T\hat{\F}\db X\right]_{\s=0}^{\s=\pi}
-\frac{1}{2}\int d\s \hat{H}_{ijk}X^{'i}\db X^j \db X^k,
\ee
where
$\hat{H}_{ijk}=\del_i\hat{\F}_{jk}+
\del_j\hat{\F}_{ki}+\del_k\hat{\F}_{ij}$.
Note that in the large $B$ limit
$\hat{H}$ is just $(1/2\pi\a')$ times $H=dB$ induced on the D-brane.

\section{$\hat{H}=0$ and Fuzzy Sphere} \label{H0}

While $\hat{H}$ maybe nontrivial in spacetime,
as long as its projection onto the D-brane vanishes,
the second term in (\ref{Om}) vanishes,
and the Poisson bracket $(\cdot, \cdot)$ for the endpoints of
the open string at $\s=0$ is
\be \label{XX}
(X_i, X_j)=2\pi\a'i\th,
\ee
where $\th=\hat{\F}^{-1}$.
The relation for the other endpoint differs only by a sign.

To quantize this system we need to replace
the Poisson brackets $(\cdot,\cdot)$
by commutators $[\cdot,\cdot]$,
but it requires some operator ordering
such that the Jacobi identity is satisfied.
We will only be concerned with the Poisson bracket in this paper.

An example is provided by the spherical D2-brane in $S^3$,
where the metric of $S^3$ is
\be
ds^{2}=k\alpha ^{\prime }\left[ d\psi ^{2}+\mathrm{sin}^{2}\psi
(d\theta^{2}+\mathrm{sin}^{2}\theta \;d\phi ^{2})\right]
\ee
and the field strength for the two-form NS-NS $B$-field is
\be
H\equiv {dB}={2k\alpha ^{\prime }}\;\mathrm{sin}^{2}\psi \;\mathrm{sin}%
\theta \;d\psi \;d\theta \;d\phi \ ,
\ee
where $k$ is an integer related to
the radius of $S^3$ by $R=\sqrt{k\a'}$.
For this $H$, we can choose $B$ to be proportional to
the volume form of the two-sphere parametrized by $(\theta ,\phi )$
on which the D2-brane wraps:
\be
B=k\alpha ^{\prime }\left( \psi -{\frac{\mathrm{sin}2\psi }{2}}\right) \;
\mathrm{sin}\theta \;d\theta \;d\phi \ . \\
\ee
The one-form field strength on the D2-brane should be \cite{BDS,Taylor}
\be
F=dA=\pi\a' n\sin\th d\th d\phi.
\ee

The energy of the D2-brane is locally minimized at
\be
\psi=\frac{\pi n}{k}
\ee
for arbitrary integer $0<n<k$ \cite{BDS,Taylor,KKZ}.
At those places,
\bea
\F=B-F=-k\a'\left(\frac{\sin 2\psi}{2}\right)\sin\th d\th d\phi.
\eea

The resulting Poisson bracket is thus
\be
(\cos\th, \phi)=-\frac{2\pi}{k}\frac{\cos\psi}{\sin\psi},
\ee
which implies that the Cartesian coordinates
satisfy the algebra of the fuzzy sphere \cite{Madore}
\be
(x_i, x_j)=\frac{2\pi}{k}\frac{\cos\psi}{\sin\psi}\eps_{ijk}x_k,
\ee
where
\be
x_1=\sin\th\cos\phi, \quad x_2=\sin\th\sin\phi,
\quad x_3=\cos\th.
\ee
In the large $k$ limit, $\psi\ll 1$,
it is $(x_i, x_j)\simeq\frac{2}{n}\eps_{ijk}x_k$.
This is in agreement with \cite{Alek,Pawel}.
For discussions on noncommutative gauge theories
on fuzzy sphere see, e.g., \cite{MSSW,Hawkins,Klimcik,Wata}.

The reason why this approximation works
is that from the flat space results we see that
the length of the open string is related to its momentum.
In the low energy limit, the momentum is very small
and so the open string is very short,
and it sees only a very small portion of the sphere,
which looks almost flat.
This also explains why the result of commutation relation
should be formally the same as the flat case.
The first main result of this paper is that
the same expressions for noncommutativity (\ref{th}), (\ref{thMB})
continue to work as long as $\hat{H}_{ijk}$ vanishes.
For the formulation of a noncommutative gauge theory
on a generic Poisson manifold see \cite{AK,JSW}.

\section{$\hat{H}\neq 0$ and New Type of Noncommutative Spaces}

What happens if $\hat{H}_{ijk}$ is not zero?
An approximate result for small $\F$ and
slow variations of $\F$ and $g$ was obtained in \cite{CHK}.
There the Jacobi identity for the algebra of $X$ and $P$
was checked to hold within the validity of this approximation.
In the following we will give a very similar derivation,
but arriving at a consistent algebra
which is valid in the low energy limit of open strings.
Our task is to find the Poisson brackets among $X$ and $P$
at $\s=0$ for the case $\hat{H}=$constant,
such that the Poisson brackets satisfy Jacobi identity to all orders,
and reduces to the previous result (\ref{th}) when $\hat{H}=0$.

We will simplify the derivation by
assuming that $X$ is linearly depending on $\sigma$.
This statement is not well defined with respect to
general coordinate transformations,
so the results we will obtain are exactly correct
only up to the first order in $X'$ or $P$,
like in a low energy approximation.

By assumption $X'$ is independent of $\s$ and
\be
X(\s)=x+\s X',
\ee
where $x$ is the coordinates of the endpoint of the string at $\s=0$.
In our convention $\s\in [0,\pi]$.
The momentum at $\s=0$ is
\be \label{P0}
p=P(\s=0)=\hat{\F}(x) X'.
\ee
From (\ref{Om}), assuming that $\del_k\hat{\F}_{ij}=$constant
so that $\hat{H}_{ijk}=$constant,
the symplectic two-form is
\bea
\Om&=&\frac{\pi}{2}(\del_k\hat{\F}_{ij}-\hat{H}_{ijk})X'^k\db
x^i\db x^j \nn \\
&&+\frac{\pi}{2}(\hat{\F}_{ij}+
\pi(\del_k\hat{\F}_{ij}-\frac{1}{2}\hat{H}_{ijk})X'^k)
(\db x^i\db X'^j-\db x^j\db X'^i) \nn \\
&&+\frac{\pi^2}{2}(\hat{\F}_{ij}+
\pi(\del_k\hat{\F}_{ij}-\frac{1}{3}\hat{H}_{ijk})X'^k)
\db X'^i\db X'^j,
\eea
where $\hat{\F}_{ij}=\hat{\F}_{ij}(x)$.
It can be explicitly checked that the symplectic two-form is closed,
so that its inverse, the Poisson bracket, satisfies the Jacobi identity.
By inverting $\Om$, we obtain the Poisson brackets for
$(x,x)$, $(x,X')$, and $(X',X')$.
To find $(x,p)$ and $(p,p)$ from these, we use (\ref{P0}).

Since it is straightforward but cumbersome to write down
the final answer of all the Poisson brackets among $x$ and $p$,
we will only write down the one involving only $x$
for the special case $\del_i\hat{\F}_{jk}=\hat{H}_{ijk}/3$,
that is,
$\hat{\F}_{ij}(x)=\hat{\F}^{(0)}_{ij}+\frac{1}{3}\hat{H}_{ijk}x^k$,
where $\hat{\F}^{(0)}_{ij}$ are constant.
The result is
\be \label{xx}
(x^i, x^j)=-2\left([I+A]^{-2}\hat{\F}^{-1}\right)^{ij},
\ee
where $I$ stands for the identity matrix and
\be
{A^i}_j=\frac{\pi}{6}\hat{\F}^{-1ik}\hat{H}_{kjm}\hat{\F}^{-1mn}p_n. \\
\ee
When $\hat{\F}^{(0)}$ is much larger than $\hat{H}$,
(\ref{xx}) is approximatly
\be
(x^i, x^j)=-2[\d^i_l-
\frac{\pi}{3}\hat{\F}^{-1ik}\hat{H}_{klm}\hat{\F}^{-1mn}p_n]
\hat{\F}^{-1lj}.
\ee
The commutation relation for the coordinates at the other
endpoint of the open string at $\s=\pi$ is the same except
a difference in sign, as it should be \cite{CH1}.
These expressions show that after quantization,
the commutator of $x$ with $x$ will in general
be a function of $x$ and $p$.

It follows that the low energy D-brane field theory lives
on a noncommutative space.
Identifying $x$ and $p$ with
the coordinates and derivatives on the D-brane,
the commutation relations among $x$ and $p$
define the differential calculus
on its noncommutative worldvolume.
The novel property that comes in when $\hat{H}\neq 0$
is that the commutator $[x^i,x^j]$ is
given by a function of $x$ and $p$,
that is, a pseudo-differential operator
on the noncommutative space.
Similarly, the commutator of $[x,p]$ and $[p,p]$
are also given by functions of $x$ and $p$,
rather than just a function of $x$.
This kind of noncommutative spaces were not considered
in the context of string theory in the pest,
but were considered long time ago \cite{Snyder,Yang}.
In \cite{CLT,BV,HL},
a similar type of noncommutative spaces (fuzzy $S^4$),
for which the commutator of coordinates is not given by
a function of the coordinates as in sec.\ref{H0}, 
were considered in matrix theory and M theory.

More care is needed to define a field theory
on such noncommutative spaces.
Since the commutator of two spacetime coordinates
generates a derivative,
how do we dinstinguish a function of $x$ only
from a function of both $x$ and $\del/\del x$
on the noncommutative space?
This problem can be solved by requiring that
a function of $x$ be written in terms of
totally symmetrized products of the $x$'s.
However, it is not clear how to define a gauge theory,
since the gauge transformation of a field,
which is a function of $x$,
will generically become a pseudo-differential operator.
On the other hand,
despite this difficulty in defining
a noncommutative gauge theory,
we should not be surprised that this kind of
noncommutative spaces appear in string theory,
since in string theory operators which are identified with
coordinates or momenta may be reinterpreted as
other physical quantities in a dual theory.

The approach used in this paper should work
even for cases in which $\hat{H}$ is not constant,
although it will be more difficult to obtain
generic expressions for the symplectic form
unless more details about $\hat{H}$ are specified.

An open membrane ending on an M5-brane in the background
of a three-form $C$ field with constant field strength
was studied in \cite{BBSS} in a limit in which
the boundary of the open membrane--a closed string--
gives a noncommutative loop algebra
on the 5-brane worldvolume.
This can also be interpreted as the noncommutativity
felt by a fundamental closed string in the background
of a constant $H$ field.
It would be interesting to see the connection between
the noncommutativities from the open and closed string
points of view.

\section*{Acknowledgment}

We thank Chong-Sun Chu and Miao Li for valuable discussions.
This work is supported in part by
the National Science Council, Taiwan, R.O.C.
and the Center for Theoretical Physics
at National Taiwan University.


\vskip .8cm
\baselineskip 22pt
\begin{thebibliography}{99}
\itemsep 0pt

\bibitem{Witten}
E. Witten:
``Noncommutative Geometry and String Field Theory'',
Nucl. Phys. B268 (1986) 253.

\bibitem{NC}
P.-M. Ho, Y.-S. Wu:
``Noncommutative Geometry and D-Branes'',
Phys. Lett. B398 (1997) 52,
hep-th/9611233;
P.-M. Ho, Y.-Y. Wu, Y.-S. Wu:
``Towards a Noncommutative Geometric Approach
to Matrix Compactification'',
Phys. Rev. D58: 026006 (1998),
hep-th/9801147;
P.-M. Ho, Y.-S. Wu:
``Noncommutative Gauge Theories in Matrix Theory'',
Phys. Rev. D58: 066003 (1998),
hep-th/9801147;
P.-M. Ho, Y.-S. Wu:
``Matrix Compactification on Orientifolds'',
Phys. Rev. D60: 026002 (1999).

\bibitem{CH1}
C.-S. Chu, P.-M. Ho:
``Non-commutative Open String and D-brane'',
Nucl. Phys. B 550 (1999) 151-168.

\bibitem{CH2}
C.-S. Chu, P.-M. Ho:
``Constrained Qunatization of Open String in Background B Field
and Noncommutative D-brane'',
hep-th/9906192.

\bibitem{CDS}
A. Connes, M. R. Douglas, A. Schwarz:
``Noncommutative Geometry and Matrix Theory:
Compactification on Tori'',
J. High Energy Phys. 02 (1998) 003, hep-th/9711162.

\bibitem{DH}
M. R. Douglas, C. Hull:
``D-Branes and Noncommutative Torus'',
J. High Energy Phys. 02 (1998) 008, hep-th/9711165.

\bibitem{Ho}
P.-M. Ho:
``Twisted Bundle on Quantum Torus
and BPS States in Matrix Theory'',
Phys. Lett. B434 (1998) 41,
hep-th/9803166.

\bibitem{BMZ}
D. Brace, B. Morariu, B. Zumino:
``Dualities of the Matrix Model From
T Duality of the Type \II String'',
Nucl. Phys. B545 (1999) 192,
hep-th/9810099.

\bibitem{CHL}
C.-S. Chu, P.-M. Ho, M. Li:
``Matrix Theory in a Constant C Field Background'',
hep-th/9911153.

\bibitem{ASS}
F. Ardalan, H. Arfaei, M. M. Sheikh-Jabbari:
``Mixed Branes and Matrix Theory on Noncommutative Torus'',
hep-th/9803067;
``Noncommutative Geometry from Strings and Branes'',
hep-th/9810072.

\bibitem{SW}
N. Seiberg, E. Witten:
``String Theory and Noncommutative Geometry'',
JHEP 9909 (1999) 032, hep-th/9908142.

\bibitem{BS}
D. Bigatti, L. Susskind:
``Magnetic Fields, Branes and Noncommutative Geometry'',
hep-th/9908056.

\bibitem{Yin}
Z. Yin:
``A Note on Space Noncommutativity'',
Phys. Lett. B466 (1998) 234,
hep-th/9908152.

\bibitem{Schom}
V. Schomerus:
``D-Branes and Deformation Quantization'',
JHEP 9906: 030 (1999), hep-th/9903205.

\bibitem{G-C}
H. Garcia-Compean, J. F. Plebanski:
``D-branes on Group Manifolds and Deformation Quantization'',
hep-th/9907183.

\bibitem{Alek}
A. Yu. Alekseev, A. Recknagel, V. Schomerus:
``Non-commutative World-volume Geometries:
Branes on $SU(2)$ and Fuzzy Spheres'',
hep-th/9908040;
%A. Yu. Alekseev, V. Schomerus, T. Strobl:
%``Closed Constraint Algebras and Path Integrals for Loop Group Actions'',
%hep-th/0001141; \\
%A. Yu. Alekseev, A. Recknagel, V. Schomerus:
``Brane Dynamics in Background Fluxes and Noncommutative Geometry'',
hep-th/0003187.

\bibitem{BDS}
C. Bachas, M. Douglas, C. Schweigert:
``Flux Stabilization of D-branes'',
hep-th/0003037.

\bibitem{Taylor}
W. Taylor:
``D2-branes in B Fields'', hep-th/0004141.

\bibitem{KKZ}
A. Kling, M. Kreuzer, J.-G. Zhou:
``$SU(2)$ WZW D-Branes and Quantized Worldvolume $U(1)$ Flux on $S^2$'',
hep-th/0005148.

\bibitem{Madore}
J. Madore, An Introduction to Noncommutative Differential Geometry
and Its Physical Applications, Cambridge U. Press, 2nd (1999).

\bibitem{Pawel}
J. Pawelczyk:
``$SU(2)$ WZW D-Branes and Their Noncommutative Geometry
from DBI Action'',
hep-th/0003057.

\bibitem{MSSW}
J. Madore, S. Schraml, P. Schupp, J. Wess:
``Gauge Theory on Noncommutative Spaces'',
hep-th/0001203.

\bibitem{Hawkins}
E. Hawkins:
``Quantization of Equivariant Vector Bundles'',
q-alg/9708030.

\bibitem{Klimcik}
C. Klim\v{c}ik:
``Gauge Theories on the Noncommutative Sphere'',
hep-th/9710153.

\bibitem{Wata}
U. Carow-Watamura, S. Watamura:
``Noncommutative Geometry and Gauge Theory on Fuzzy Sphere'',
hep-th/9801195.

\bibitem{AK}
T. Asakawa, I. Kishimoto:
``Noncommutative Gauge Theories from Deformation Quantization'',
hep-th/0002138.

\bibitem{JSW}
B. Jur\v{c}o, P. Schupp, J. Wess:
``Noncommutative Gauge Theory for Poisson Manifolds'',
hep-th/0005005.

\bibitem{CHK}
C.-S. Chu, P.-M. Ho, Y.-C. Kao:
``Worldvoulme Uncertainty Relations for D-Branes'',
hep-th/9904133.

\bibitem{Snyder}
H. S. Snyder:
``Quantized Space-Time'', Phys. Rev. 71 (1946) 38;
``The Electromagnetic Field in Quantized Space-Time'',
Phys. Rev. 72 (1947) 68.

\bibitem{Yang}
C. N. Yang:
``On Quantized Space-Time'',
Phys. Rev. 72 (1947) 874.

\bibitem{CLT}
J. Castelino, S. Lee, W. Taylor:
``Longitudinal 5-Branes as 4-Sphere in Matrix Theory'',
Nucl. Phys. B526 (1998) 334, hep-th/9712105.

\bibitem{BV}
M. Berkooz, H. Verlinde:
``Matrix Theory, AdS/CFT and Higgs-Coulomb Equivalence'',
JHEP 9911: 037 (1999), hep-th/9907100.

\bibitem{HL}
P.-M. Ho, M. Li:
``Fuzzy Spheres in AdS/CFT Correspondence
and Holography from Noncommutativity'',
hep-th/0004072.

\bibitem{BBSS}
E. Bergshoeff, D. S. Berman, J. P. van der Schaar, P. Sundell:
``A Noncommutative M-Theory Five-Brane'',
hep-th/0005026.

\end{thebibliography}
\end{document}